# Quasi-Two-Dimensional Magnon Identification in Antiferromagnetic FePS$_3$ via Magneto-Raman Spectroscopy

*Amber McCreary,[1] Jeffrey R. Simpson,[1,2] Thuc T. Mai,[1,3] Robert D. McMichael,[1] Jason E. Douglas,[4] Nicholas Butch,[5] Cindi Dennis,[4] Rolando Valdes Aguilar,[3] and Angela R. Hight Walker[1,*]*

[1]Nanoscale Device Characterization Division, Physical Measurement Laboratory, National Institute of Standards and Technology, Gaithersburg, Maryland 20899, USA

[2]Department of Physics, Astronomy, and Geosciences, Towson University, Towson, MD 21252, USA

[3]Department of Physics, The Ohio State University, Columbus, OH 43210, USA

[4]Materials Science and Engineering Division, Material Measurement Laboratory, National Institute of Standards and Technology, Gaithersburg, Maryland 20899, USA

[5]NIST Center for Neutron Research, National Institute of Standards and Technology, Gaithersburg, Maryland 20899, USA

*Email: angela.hightwalker@nist.gov

**ABSTRACT:**

Recently it was discovered that van der Waals-bonded magnetic materials retain long range magnetic ordering down to a single layer, opening many avenues in fundamental physics and potential applications of these fascinating materials. One such material is $FePS_3$, a large spin (S=2) Mott insulator where the Fe atoms form a honeycomb lattice. In the bulk, $FePS_3$ has been shown to be a quasi-two-dimensional-Ising antiferromagnet, with additional features in the Raman spectra emerging below the Néel temperature ($T_N$) of approximately 120 K. Using magneto-Raman spectroscopy as an optical probe of magnetic structure, we show that one of these Raman-active modes in the magnetically ordered state is actually a magnon with a frequency of ≈3.7 THz (122 cm$^{-1}$). Contrary to previous work, which interpreted this feature as a phonon, our Raman data shows the expected frequency shifting and splitting of the magnon as a function of temperature and magnetic field, respectively, where we determine the g-factor to be ≈2. In addition, the symmetry behavior of the magnon is studied by polarization-dependent Raman spectroscopy and explained using the magnetic point group of $FePS_3$.

## I. INTRODUCTION

Since the isolation of monolayer graphene in 2004,[1] there has been a surge of research into van der Waals layered materials, where the strong intralayer and weak interlayer coupling allows for isolation of layers that are only a few atoms thick. These materials exhibit a wide range of electronic properties, including semiconducting, metallic, insulating, superconducting, and charge density waves,[2-4] allowing for device architectures composed of solely two-dimensional (2D) materials. While significant research has been dedicated thus far to studying the optical, mechanical, and electrical properties of 2D materials,[3,5-9] exploring magnetism is still in its infancy, even though 2D magnetic materials provide a solid-state platform to experimentally access fundamental, low-dimensional physics.[10,11] Additionally, any 2D magnetic material would likely still possess the captivating properties of 2D materials, including extremely large mechanical flexibility,[12,13] efficient tuning of transport properties with an electric field,[14-17] relative ease of chemical modification,[18,19] as well as the ability to create van der Waals stacked heterostructures.[20] These myriad of tuning parameters could unlock opportunities for custom-engineered magnetoelectric and magneto-optical devices, where 2D magnets coupled with other technologically relevant materials could realize unprecedented capabilities in fields such as spintronics.[11,21,22]

Until recently, long range magnetic order was believed to be impossible in 2D systems due to thermal fluctuations.[23] In early 2017, however, intrinsic ferromagnetism was observed down to the few-layer and monolayer limit in two different, layered materials with magnetic anisotropy, including $Cr_2Ge_2Te_6$ and $CrI_3$.[24,25] For $CrI_3$, it was also shown that the interlayer magnetic ordering (*i.e.* ferromagnetically or antiferromagnetically coupled) was dependent on the number of layers,[25] and could be controlled by an external electric field.[14,26] Transition metal phosphorus trisulfides $XPS_3$ ($X$ = Fe, Mn, Ni, *etc*.) are another class of van der Waals antiferromagnets that are being studied in the 2D limit. Interestingly, although $FePS_3$ ($T_N \approx 120$ K),[27] $MnPS_3$ ($T_N \approx 78$ K),[27] and $NiPS_3$ ($T_N \approx 155$ K)[27] are isostructural, they have different spin structures below the Néel temperature. The choice of transition metal results in varied magnetic phenomena, since the spins align antiferromagnetically in different fashions, including Néel, zigzag, or stripe ordering.[28,29] $FePS_3$, which is a Mott insulator,[30,31] is especially intriguing as a 2D Ising antiferromagnet on a honeycomb lattice.[27,32] In addition, long-distance magnon transport (several micrometers) has been experimentally observed in MnPS$_3$, demonstrating that these materials are viable candidates for future magnonic devices.[22]

Raman spectroscopy, being non-destructive and highly sensitive to minute perturbations, is a powerful technique to study various properties of quantum materials, including the effects of layer number,[33,34] strain,[35] defects/doping,[36] electron-phonon coupling,[37,38] phase transitions,[39] spin-phonon coupling,[40] and magnetic excitations.[41,42] In addition, unlike other measurements that require bulk, large-area crystals, such as neutron diffraction, X-ray diffraction, or magnetic susceptibility, Raman spectroscopy can probe atomically thin flakes with diffraction-limited spatial resolution. The Raman spectra of bulk *X*PS$_3$ materials has been studied since the 1980's,[28,43,44] and only recently extended to samples in the monolayer limit for NiPS$_3$[45] and FePS$_3$.[46,47] In particular, FePS$_3$ is an interesting candidate to study using Raman spectroscopy because, due to the antiferromagnetic alignment of the spins and the resulting increase in the unit cell, Brillouin zone folding leads to new modes appearing below $T_\text{N}$.[46,47]

In this work, we show that one of the modes that appears below $T_\text{N}$, which has been recently attributed to a phonon,[46] is actually a quantized spin wave, *i.e.* a magnon. This antiferromagnetic magnon, with a frequency of 3.7 THz ($\approx$122 cm$^{-1}$), softens with temperature more rapidly than typical phonon modes and splits upon application of a magnetic field, as expected for antiferromagnetic magnons. The frequency of the non-degenerate magnons depends linearly on magnetic field, with an effective g-factor $g \approx 2$. However, its symmetry behavior with respect to the polarization vectors of the incoming and scattered light contradicts the long accepted work of Fleury and Loudon which states that Raman scattering of first-order magnon excitations involves antisymmetric Raman tensors and can thus only be observed in cross-polarized light configurations.[41] Instead, we observe the magnon in FePS$_3$ in both parallel and cross configurations, thus showing that the established magnon symmetry rule lacks the generality that has been suggested by previous literature. We explain the symmetry of the observed magnon using the magnetic point group with the inclusion of complex tensor elements. Because of the small interlayer exchange coupling, and thus quasi-2D magnetic nature of bulk FePS$_3$, the magnon observed herein is also expected to be quasi-2D. To the best of our knowledge, this work represents the first verification of a magnon in a quasi-2D magnet using magneto-Raman spectroscopy. Furthermore, this work highlights temperature-dependent, magneto-Raman spectroscopy as an important technique to explore properties of magnetic excitations, such as frequencies and lifetimes, and to aid in investigating next-generation, magnonic devices. FePS$_3$ may prove a better candidate for magnon transport than the previously studied MnPS$_3$ as the magnon frequency is an order of magnitude higher, promising faster switching capabilities.[22,48]

## II. RESULTS AND DISCUSSIONS

**A. Initial Characterization: Sample Details and Magnetization** Large crystals of FePS$_3$ were mechanically exfoliated using adhesive tape[49] onto 300 nm SiO$_2$ thermally grown on Si(100). Thicker flakes were selected for Raman spectroscopy measurements due to the more intense scattering signal. Bulk FePS$_3$ crystallizes in the monoclinic structure with the space group C2/m.[50] In the *a-b* plane, as is shown in Figure 1a, the Fe atoms form a honeycomb lattice. In the center of the Fe honeycomb lattice is a [P$_2$S$_6$]$^{4-}$ unit, where the P$_2$ dimers orient normal to the surface and coordinate with six S atoms. The magnetic structure has been a source of controversy,[32,50-53] where in both proposed structures, the moments of the large spin Fe$^{2+}$ ions (S=2) are aligned normal to the *a-b* plane and ordered in ferromagnetic chains that couple antiferromagnetically. The proposed structures differed from each other, however, by a 120° rotation of the ferromagnetic chains, as well as whether the layers themselves are ferromagnetically or antiferromagnetically coupled. Recent neutron scattering by D. Lançon *et al.*[53] of large, single crystals of FePS$_3$ confirmed the magnetic structure consists of ferromagnetic chains oriented along the *a*-axis, with antiferromagnetically coupled layers along the *c*-axis, as shown in Figures 1b and 1c.

Using a SQUID magnetometer, the magnetization of the FePS$_3$ crystals (*i.e.* before exfoliation) was measured both parallel and perpendicular to the *a-b* plane with an applied magnetic field of 0.1 T. The magnetization versus temperature from 5 K to 300 K is shown in Figure 1d. Similar to earlier reports,[27,46,47] the strong anisotropy in FePS$_3$ results in higher magnetization perpendicular to the *a-b* plane, which is along the spin direction in the antiferromagnetic state. In a 2D antiferromagnet, $T_N$ is defined to be the temperature where there is a maximum in the slope of the magnetization vs. temperature curve.[27] Thus, by taking the first derivative of the magnetization with respect to temperature (Figure 1e), $T_N$ is found to be ≈118 K. The broad maximum above $T_N$ in the magnetization vs. temperature for both parallel and perpendicular orientations is typical for low-dimensional magnetic systems and has been attributed to short-range spin-spin correlation.[27]

**B. Raman Spectroscopy and Emergence of New Modes** The majority of the peaks in the Raman spectrum (Figure 1f) can be ascribed to phonons. From the symmetry of bulk FePS$_3$, there are 30 zone-center phonons with irreducible representations: $\Gamma = 8A_g + 6A_u + 7B_g + 9B_u$, of which only the $A_g$ and $B_g$ modes are Raman active. Figure 1f shows the Raman spectra of bulk FePS$_3$ (laser excitation wavelength λ = 514.5 nm) in the 180° back-scattering geometry at T = 135 K and T = 15 K, which corresponds to above and well below the Néel temperature, respectively. Phonons from the $\Gamma$ point ($\omega > 150$ cm$^{-1}$) are labeled by whether they are an $A_g$ mode or a combination of $A_g/B_g$ modes based on previous first principle calculations as well as their measured orientation-dependent frequency since a peak that is a combination of $A_g/B_g$ modes will appear to shift in frequency as the light polarization is rotated with respect to the crystal orientation.[47,54] The phonon modes $A_g^2/B_g^2$, $A_g^3$, $A_g^4/B_g^4$, and $A_g^5$ are mostly ascribed to vibrations of the $(P_2S_6)^{4-}$ unit, and have similar frequencies for the different $XPS_3$ materials due to the lack of contribution from the metal atom.[54] While the $A_g$ and $A_g/B_g$ modes in the spectra are similar in terms of frequency and intensity when comparing above and below the Néel temperature, significant changes are seen in the modes below 150 cm$^{-1}$. Above the Néel temperature, there is a broad, asymmetric peak "$\Psi$" that exists even up to room temperature. At T = 15 K, however, there are four new modes that appear in this frequency range, labeled: $\Psi_1$, $\Psi_2$, $\Psi_3$, and $\Psi_4$.

The evolution of the Raman spectra of mode $\Psi$ into four modes below $T_N$ has been observed since the 1980's,[28,43,44] yet there is still uncertainty regarding the nature of the transition and the origin of each component. It is well known that antiferromagnetic ordering in FePS$_3$ leads to a doubling of the unit cell and subsequent zone folding, such that the M point in the paramagnetic Brillouin zone folds into the $\Gamma$ point of the antiferromagnetically ordered Brillouin zone. From the phonon dispersion calculated by X. Wang *et al.*,[47] $\Psi_1$ and $\Psi_2$ appear to be M point phonons present below the Néel temperature due to magnetic-ordering induced zone folding. X. Wang *et al.* also state that $\Psi_3$ is a K point phonon, however they do not indicate how it becomes Raman-active since zone folding does not cause the K point to fold into the $\Gamma$ point. On the contrary, J.-U. Lee *et al.*[46] attribute these modes to very different origins. From density functional theory calculations, they suggest $\Psi_2$ is a $\Gamma$ point phonon without magnetic ordering, while $\psi_3$ and $\psi_4$ have both $\Gamma$ and M point components, and would thus appear above $T_N$. Other older publications also observed $\psi_1$-$\psi_4$, but only comment on possible origins of $\psi_1$ and $\psi_2$.[44,55] Thus, although the emergence of these modes can provide insight on the transition to a magnetically ordered state in FePS$_3$, there is still substantial debate as to the source of each mode. In this work, we shed light on the origin of $\psi_4$, where we unequivocally show that it is a magnon.

**C. Temperature Dependence of Raman Modes:** Figure 2a shows the temperature dependence of the Raman spectra from 110 K down to 40 K in steps of 10 K. $\psi_4$ is not observed until temperatures below 100 K, well below $T_N$ of ≈118 K. As the temperature is further decreased, $\psi_4$ displays a dramatic blueshift (higher in frequency) with temperature, especially compared with the other modes, including $\psi_1$, $\psi_2$, $\psi_3$ and the $\Gamma$-point phonons. This effect is illustrated in Figure 2b, which shows the frequencies of the different modes relative to their frequency at 15 K ($\omega - \omega_{15K}$), including standard errors from fitting each peak with a Voigt function. While the frequency changes of the other modes only range from 0.01% to 0.57% (1 cm$^{-1}$ to 2 cm$^{-1}$ shifts), as is typical for anharmonic lattice effects, the frequency of $\psi_4$ changes by as much as 6.2% (8 cm$^{-1}$ shift) as the temperature is increased to 100 K. The strong shift in frequency with respect to temperature offers the first indication that this mode is attributed to a magnon, as suggested by T. Sekine *et al.* in 1990 for FePS$_3$.[56] Similar magnon temperature dependence has been reported for other antiferromagnets, such as FeF$_2$,[57,58] NiF$_2$,[59] and MnF$_2$.[60] The assignment of this mode to a magnon is further supported by recent inelastic neutron scattering measurements that observed a $\Gamma$ point magnon at ≈15.1 meV (122 cm$^{-1}$),[53,61] which coincides in frequency to our observed mode $\psi_4$.

**D. Magnons and Magneto-Raman Spectroscopy of FePS$_3$** Due to the contrasting behavior of phonons and magnons in an applied magnetic field, we utilize our unique magneto-Raman capabilities to investigate the magnetic field dependence of the modes in FePS$_3$ to identify if they have magnetic character. We detail our findings from the magnetic-field dependent Raman spectra of FePS$_3$. Then, by discussing how magnons behave under applied magnetic fields, we assign $\Psi_4$ as a Raman-active magnon.

An exfoliated flake of bulk FePS$_3$ was cooled to T = 5 K, after which a magnetic field was applied normal to the *a-b* plane, *i.e.* parallel/antiparallel to the directions of the spins in the antiferromagnetically ordered state. As seen in Figure 3a, when the magnetic field is increased from 0 T to 9 T, $\Psi_4$ splits into two modes, labeled as $\Psi_4^{(1)}$ and $\Psi_4^{(2)}$, and the splitting increases with the applied magnetic field. Figure 3b shows the frequencies of $\Psi_4^{(1)}$ and $\Psi_4^{(2)}$ as a function of magnetic field, where the splitting is symmetric for positive and negative magnetic fields. From a linear fit, the extracted slopes for the $\Psi_4^{(1)}$ and $\Psi_4^{(2)}$ branches are 0.93 ± 0.02 cm$^{-1}$/T and 0.94 ± 0.01 cm$^{-1}$/T, respectively. This splitting only occurs for $\Psi_4$, where the other $\Psi$ modes and phonon modes remain as single peaks. As discussed below, this behavior matches well with the expected magnetic field dependence of magnons.

Spin waves, which as quasiparticles are called magnons, are collective excitations of the spins in magnetic materials. The classical dynamics of the magnetization at the Γ-point in an antiferromagnet can be modeled following F. Keffer and C. Kittel[62] as two interacting macrospins (Figure 3c) representing the spin-up sublattice magnetization (M$_1$, pink) and spin-down sublattice magnetization (M$_2$, green) that orient in the +*z* and -*z* directions in equilibrium, respectively. The excitation of a magnon causes the magnetization on both sublattices to precess about their equilibrium directions. Figures 3(d) and 3(e) show side and top view illustrations of the two normal modes for $M_1$ and $M_2$ precession. The frequencies of the normal modes are given by:

$$\omega_{k=0} = \gamma\{(2H_E + H_A)H_A\}^{1/2} \pm \gamma H_0 \qquad (1)$$

where $H_E$ is the exchange field, $H_A$ is the anisotropy field, $H_0$ is an applied field, and $\gamma = \frac{g\mu_B\mu_0}{2\pi\hbar}$ is the gyromagnetic ratio, which for free electrons ($g \approx 2.0023$)[63] is approximately 28.025 GHz/T, or 0.9348 cm$^{-1}$/T. Without an applied magnetic field ($H_0 = 0$), the frequencies of the magnon normal modes are degenerate ($E = E_0$). However, an applied magnetic field $H_0$ along $z$ will cause a Zeeman splitting, leading to two separate modes with $E < E_0$ (Figure 3d) and $E > E_0$ (Figure 3e), both changing in magnetic field with a slope of $|\gamma|$. From Equation 1, a splitting of the mode with applied magnetic field offers a conclusive method to determine if a Raman mode is a magnon, as a magnetic field dependence is not expected for a typical phonon.

The behavior of the normal modes of magnons in a magnetic field described above matches with the observed magnetic-field dependence of $\psi_4$, where the application of a magnetic field parallel to the spin direction leads to a lifting of the degeneracy of the magnon frequency into two components, $\psi_4^{(1)}$ and $\psi_4^{(2)}$. The frequencies of the two non-degenerate magnon modes obey a linear response as a function of magnetic field, in agreement with Equation 1, with the two slopes of equal (within error) to the predicted slope $\gamma = 0.93481$ cm$^{-1}$/T for the free electron limit of $g \approx 2.0023$. Using our experimental values of $\gamma$ from the slopes of the linear fits in Figure 3b, we estimate that for bulk FePS$_3$, the effective magnon g-factor is $g \approx 1.99 \pm 0.05$. Thus, through magneto-Raman spectroscopy, we can unequivocally assign the mode $\psi_4$ at $\approx$122 cm$^{-1}$ as a magnon. From the full-width at half maximum (FWHM) value of approximately 3 cm$^{-1}$, which includes instrument broadening, we estimate the lower bound of the magnon lifetime to be on the order of 10 picoseconds.

**E. Magnon Symmetry Behavior and Magnetic Point Group** Raman scattering from magnons in both ferromagnets and antiferromagnets has been theoretically discussed by numerous authors, with one of the most cited works by P.A. Fleury and R. Loudon from 1968.[41] In their work, the resulting polarization selection rules of Raman scattering from first order magnon excitations is purely antisymmetric with respect to the polarization of the incoming ($\varepsilon_i$) and scattered ($\varepsilon_s$) light. Thus, from their theory, a magnon mode can be observed only in the perpendicular (or crossed) polarization configurations ($\varepsilon_i = \varepsilon_s \pm 90°$), regardless of the orientation with respect to the crystallographic axes. This theory has worked well to describe a variety of ferromagnetic and antiferromagnetic materials with Raman-observed magnons, including $FeF_2$,[41] $MnF_2$,[41] $YFeO_3$,[64] and $Cd_{1-x}Mn_xTe$.[65]

To probe the symmetry behavior of the magnon in $FePS_3$, $\varepsilon_i$ and $\varepsilon_s$ were selected as shown in the inset of Figure 4a, where incoming light $\varepsilon_i$ is at an arbitrary angle φ with respect to the crystallographic *b*-axis and the scattered light $\varepsilon_s$ is at a controlled angle θ with respect to $\varepsilon_i$. Figure 4a shows that the magnon in $FePS_3$, can be observed in parallel polarization (θ=0°) in addition to perpendicular polarization (θ=90°). When θ is varied between 0° and 360°, a two-fold symmetric pattern emerges where the magnon never disappears, but instead has a minimum at θ=60° and maximum at θ=150°, although these angle values are dependent on φ.

The theory presented in Fleury and Loudon applies rigorously to magnetic ions with orbital angular momentum ground states of L=0, or those with crystal-field quenching of the ground state to L=0,[41] as is the case for the well-studied antiferromagnetic magnon example of $FeF_2$.[66,67] For materials that do not satisfy the L=0 ground state condition, such as $FePS_3$,[68] the magnetic excitations are more difficult to treat; the general "antisymmetric" nature of the magnon Raman tensor no longer applies. Instead, we use the symmetry properties of the *magnetic* point group of bulk FePS3 to understand the selection rules and polarization dependence of the $\Psi_4$ magnon mode shown in Figure 4.

When a material orders antiferromagnetically, the symmetry generally lowers because the spin up and spin down sites are no longer equivalent. In this regard, for the magnetic point group, unitary operations are symmetry operations that preserve the spin direction. Other operations that do not preserve the spin direction may become allowed with the application of time reversal symmetry and are referred to as antiunitary operations. The magnetic point group for bulk FePS3, which considers both unitary and antiunitary operations is 2´/m, where the prime means time reversal. The Raman tensors for this magnetic group are given by the co-representations:[69]

$$DA' = \begin{pmatrix} A & B & 0 \\ D & E & 0 \\ 0 & 0 & I \end{pmatrix}, \quad DA'' = \begin{pmatrix} 0 & 0 & C \\ 0 & 0 & F \\ G & H & 0 \end{pmatrix}$$

The experimental 180° backscattering configuration incident on the honeycomb plane can access only the upper left $2 \times 2$ block of these tensors, thus eliminating consideration of $DA''$. The Raman intensity is proportional to $|\varepsilon_i \cdot DA' \cdot \varepsilon_s|^2$, where $\varepsilon_i = (\sin \varphi, \cos \varphi, 0)$ and $\varepsilon_s = (\sin \varphi + \theta, \cos \varphi + \theta, 0)$ are the incident and scatter light polarization vectors, respectively.

It should be noted that if $A$, $B$, $C$, and $E$ are purely real numbers, the intensity profile as a function of θ would necessarily have two nodes, *i.e.* two angles where the intensity goes to zero, which is contrary to the polar plot we observed in Figure 4b. These nodes, however, disappear from the $DA'$ symmetry response when the tensor elements are complex, which is the general case for absorbing materials:[70]

$$DA' = \begin{pmatrix} A & be^{i\delta_B} \\ de^{i\delta_D} & fe^{i\delta_F} \end{pmatrix}$$

where $\delta_B$, $\delta_D$, and $\delta_F$ are the complex phase factors of the real tensor elements $B$, $D$, and $F$ with respect to $A$. Considering the complex tensor elements, there are multiple combinations of the amplitude and phase factors that reproduce the polar plot in Figure 4b. However, there are two points that remain clear. First, at least one of the phase factors must be non-zero in $FePS_3$ to agree with the non-nodal nature of the experimental polar plot. Second, the symmetry response of the magnon $M$ in $FePS_3$ deviates strongly from the antisymmetric nature predicted by Fleury and Loudon[41] for magnetic ions with L=0 ground state. Thus, there is no reason, *a-priori*, to expect that a magnon will appear only in cross-polarization configurations for these honeycomb-like structures, such as $FePS_3$, α-$RuCl_3$, or $CrI_3$,[71] without considering the magnetic point group symmetry and associated Raman tensors.

In principle, one should be able to use the generators of rotations (*i.e.* angular momenta $J_x, J_y$, and $J_z$) generally given in the character table as guides for the magnon symmetry behavior. For example, in $FePS_3$ where the spins are oriented along the *z*-direction (normal to *a-b* plane), the excitation of a magnon reduces the magnitude of the magnetization $M_z$, which is then converted to the precession of the spins in the *x-y* (*a-b*) plane, $M_x$ and $M_y$, as depicted in Figures 3d and 3e. Thus, a magnon in $FePS_3$, which is a magnetization rotation in the *x-y* plane, can only have the same symmetry as $J_x$ and $J_y$, which are generators of rotations. While the character tables that state these transformations are calculated for the non-magnetic space groups,[72-74] we were unable to find a reference that tabulated these for the magnetic space groups. Deriving such transformations is out beyond the scope of this work.

## III. CONCLUSIONS

In conclusion, we examined the temperature- and magnetic field-dependent Raman spectra of bulk $FePS_3$, where new Raman-active modes appear in the antiferromagnetically ordered state. One of these modes ($\psi_4$), with a frequency of ≈122 cm$^{-1}$ (15.1 meV), was previously assigned as a phonon mode appearing due to zone folding. However, the strong temperature shift (compared to other phonon modes) and splitting (and linear shifting) of the mode with applied magnetic field observed herein allows this mode to be unequivocally assigned as a magnon. We also investigated the magnon symmetry behavior, which is not purely antisymmetric, and explained its behavior using the magnetic point group of $FePS_3$ (2´/m). This work will aid in future studies of magnetic excitations in similar magnetic layered materials, such as α-$RuCl_3$ and $CrI_3$, and demonstrates temperature- and magnetic field-dependent Raman spectroscopy as a technique to probe magnon phenomena in 2D materials, including for possible magnon transport applications. The quasi-2D magnetic nature of bulk $FePS_3$, with weak interlayer exchange coupling, indicates that the magnon in bulk $FePS_3$ is also quasi-2D. To the best of our knowledge, this is the first verification of a quasi-2D magnon in a layered material using magneto-Raman spectroscopy.

## IV. EXPERIMENTAL DETAILS

Bulk $FePS_3$ crystals were purchased from 2D Semiconductors[†] and then exfoliated via the adhesive tape onto 300 nm $SiO_2$ thermally grown on Si(100).[49] Raman spectra were measured with the 514.5308 nm excitation wavelength of an Argon ion laser in the 180° backscattering configuration using a triple grating Raman spectrometer (Horiba JY T64000[†], 1800 $mm^{-1}$ grating) coupled to a liquid nitrogen cooled CCD detector. Polarization was selected and controlled using ultra broadband polarizers and achromatic half wave plates. To perform temperature- and magnetic-field dependent Raman, the sample was placed into an attoDRY1000 cryostat (Attocube Inc.[†]), where the sample holder is pumped to ≈1x$10^{-3}$ Pa (≈7x$10^{-6}$ Torr), backfilled with helium gas, and cooled. Micrometer-sized flakes were studied by focusing the laser with a white light camera onto the sample with a low-temperature, magnetic field compatible objective (50×, N.A. 0.82) and *xyz* nano-positioners. Integration times were approximately 12 minutes and the laser power was kept below 300 µW to reduce local heating of the sample. Spectra with applied magnetic field were corrected for Faraday rotation in the objective using half wave plates external to the cryostat. Magnetization vs. temperature measurements were performed on a Superconducting Quantum Interference Device-based Vibrating Sample Magnetometer (SQUID-VSM; Quantum Design, Inc.[†]). The piece of the unexfoliated $FePS_3$ crystal was mounted on to a quartz holder using GE varnish (LakeShore Cryotronics, Inc.[†]), and then measured under vacuum (< 1 kPa). The magnetization was measured under an applied magnetic field $\mu_0 H = 0.1$ T from 5 K to 300 K and back down to 5 K. The temperature was varied in steps of 5 K between 5 K – 300 K, 1 K between 30 K – 150 K, and then 5 K between 150 K – 300 K.

[†]Certain commercial equipment, instruments, or materials are identified in this manuscript in order to specify the experimental procedure adequately. Such identification is not intended to imply recommendation or endorsement by the National Institute of Standards and Technology, nor is it intended to imply that the materials or equipment are necessarily the best available for the purpose.

## V. ACKNOWLEDGEMENTS

A.M., T.T.M., and A.R.H.W. would like to acknowledge the National Institute of Standards and Technology (NIST)/National Research Council Postdoctoral Research Associateship Program and NIST-STRS (Scientific and Technical Research and Services) for funding. Work at The Ohio State University was supported by the Center for Emergent Materials, an NSF MRSEC under grant DMR-1420451. We would also like to acknowledge Prof. Natalia Drichko for insightful discussions and Hans Nembach for a careful readthrough of the paper.

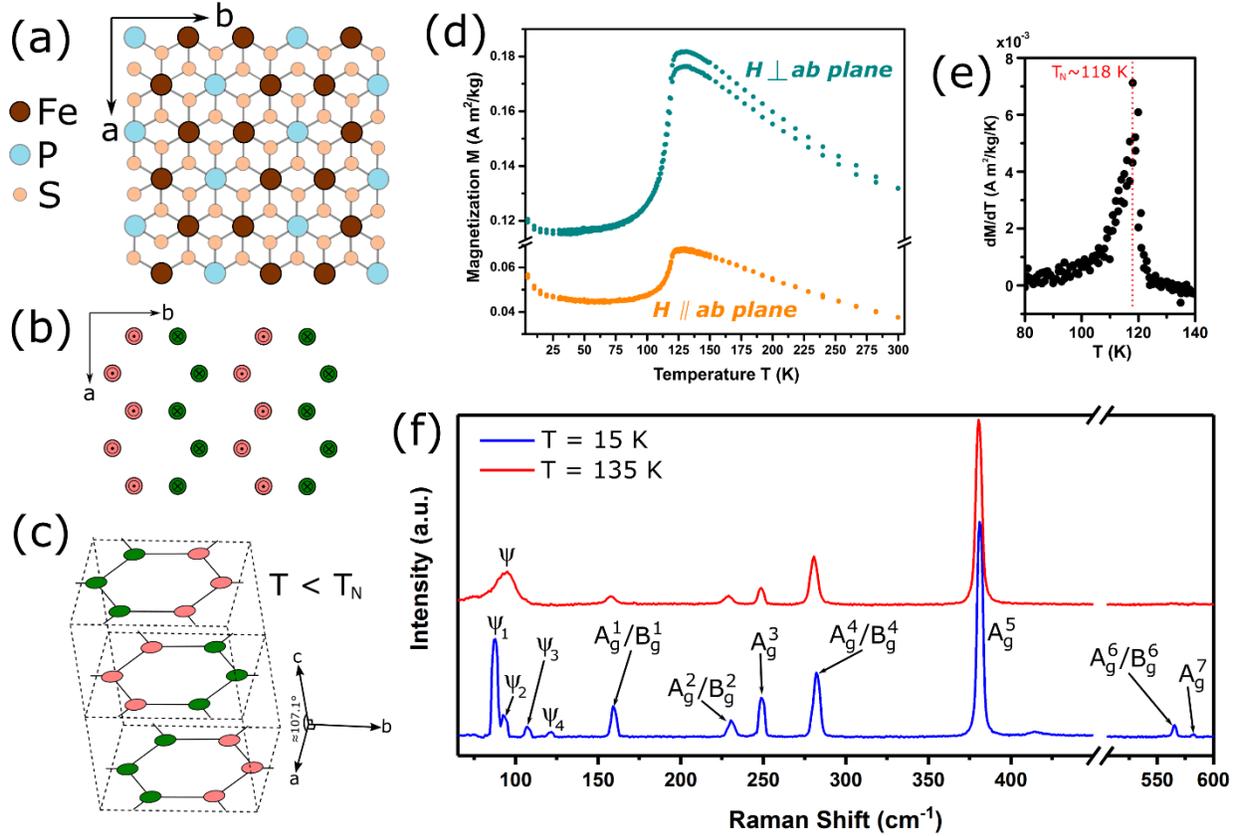

**Figure 1**. (a) Top view of the crystal structure of FePS$_3$, where the Fe atoms form a honeycomb lattice. The a- and b-axes are labeled. (b) Magnetic structure of the Fe atoms in the *a-b* plane below $T_N$. The spins are collinear and normal to the surface, where pink indicates spin "up" (pointing out of the page) and green indicates spin "down" (pointing into the page). The ferromagnetic chains are aligned along the *a*-axis.[50] (c) Below $T_N$, the layers are stacked antiferromagnetically, resulting in an increase in the unit cell along the c-axis. (d) Magnetization (*M*) vs. temperature (*T*) for a bulk crystal of FePS$_3$, where the applied magnetic field (*B* = 0.1 T) is parallel (orange) or perpendicular (cyan) to the *a-b* plane. (e) *dM/dT* of (d) with an inflection point at $T_N \approx 118$ K. (f) Raman spectra above (*T* = 135 K) and below (*T* = 15 K) $T_N$.

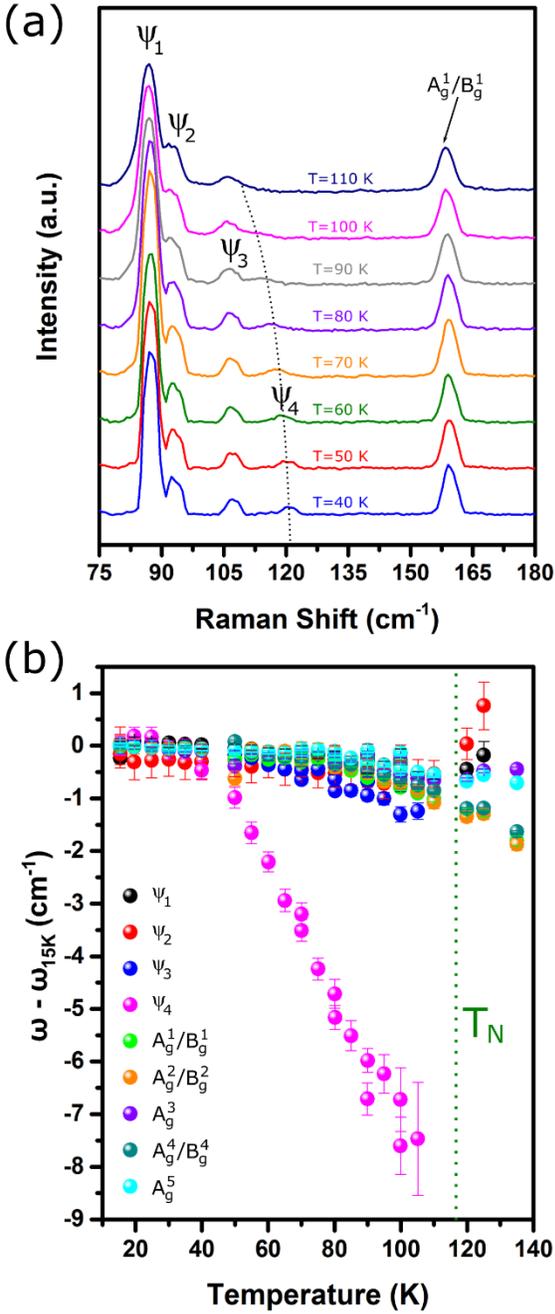

**Figure 2.** (a) Temperature-dependent Raman spectra of bulk FePS$_3$ below $T_N$. The $\Psi_4$ mode appears below 100 K and its frequency dramatically increases with decreasing temperature. (b) Frequencies of the various modes (relative to frequency at 15 K). While the other $\Psi$ modes and phonon modes only shift by 1 cm$^{-1}$ to 2 cm$^{-1}$, as is expected by lattice anharmonic effects, the frequency of $\Psi_4$ shifts by as much as 8 cm$^{-1}$ before it is no longer measurable above 100 K.

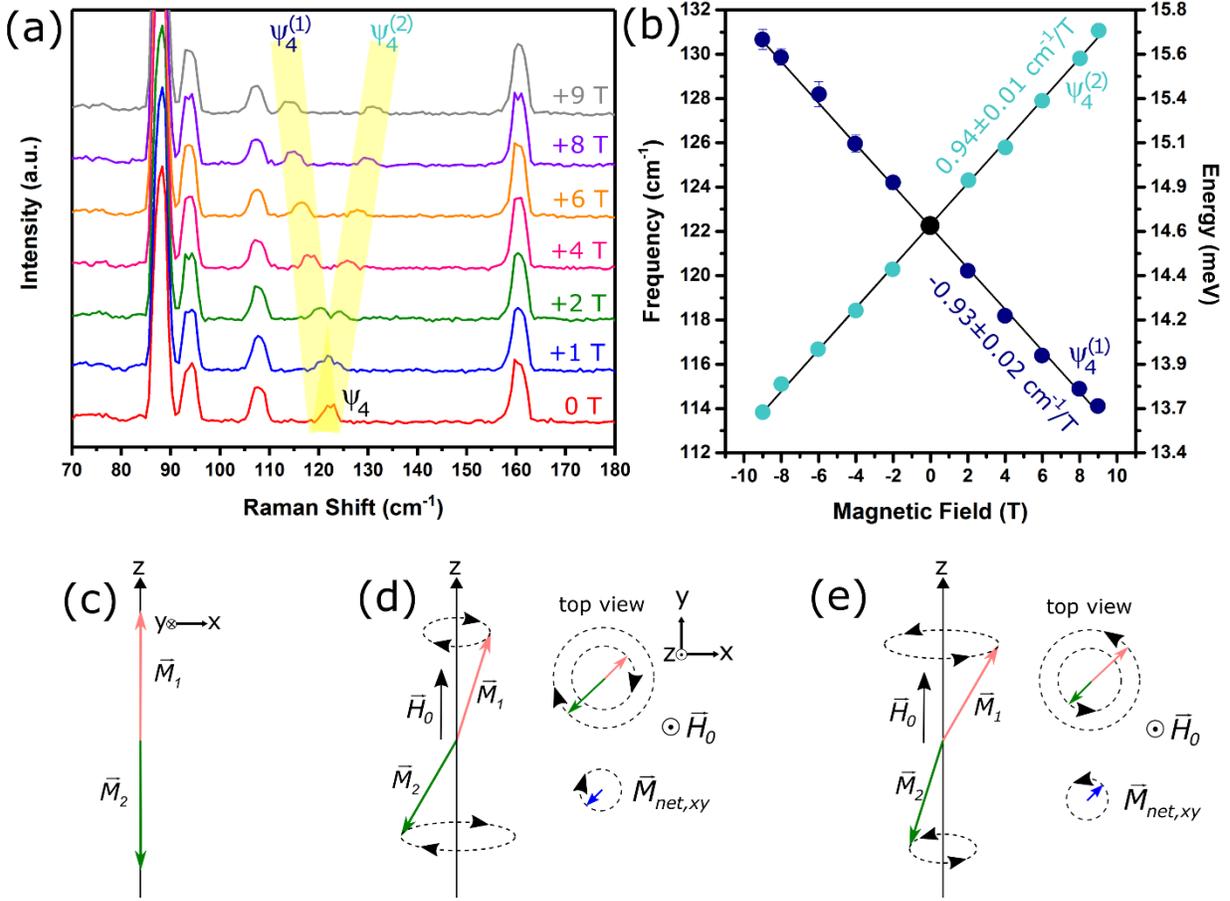

**Figure 3.** (a) Magnetic-field dependent Raman of $FePS_3$ at $T = 5K$, showing the splitting of $\Psi_4$ into two components, $\Psi_4^{(1)}$ and $\Psi_4^{(2)}$, where the frequency of $\Psi_4^{(1)}$ ($\Psi_4^{(2)}$) decreases (increases) with increasing magnetic field. (b) Frequency vs. magnetic field of $\Psi_4^{(1)}$ and $\Psi_4^{(2)}$ with the slopes of the linear fits for the two branches. (c) If we consider two magnetization sublattices $M_1$ (pink) and $M_2$ (green) in $FePS_3$, in the ground state $M_1$ and $M_2$ point in the $+z$ and $-z$ directions. (d, e) The magnon spectrum includes two normal modes with net moments Mnet in the $x,y$ plane that precess in opposite directions. The degeneracy of these modes is broken in an applied field $H_0\hat{z}$ with the energies of the modes in (d) and (e) respectively decreasing and increasing with $H_0$. In terms of this simple model, the modes in (d) and (e) would represent $\Psi_4^{(1)}$ and $\Psi_4^{(2)}$ in (b), respectively.

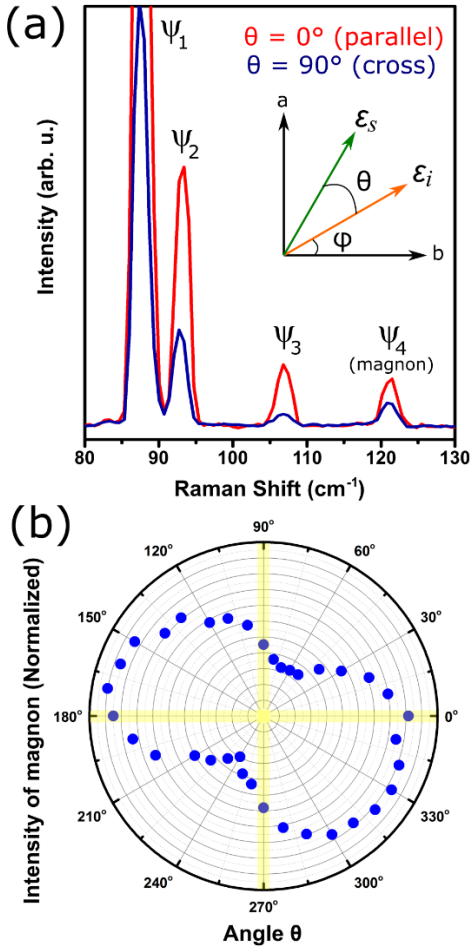

**Figure 4**. (a) Raman spectra ($T = 10K$) showing intensity of magnon $\psi_4$ in parallel ($\theta=0°$) and cross ($\theta=90°$) polarization configurations at some arbitrary angle $\varphi$. Inset shows a schematic defining angles $\theta$ and $\varphi$ with respect to the incoming ($\varepsilon_i$) and scattered ($\varepsilon_s$) light polarization and the *a*- and *b*- crystal axes. (b) Polar plot of intensity of the magnon as a function of $\theta$ at a constant angle $\varphi$. The intensity scale is represented by the radial lines that span from 0.1 to 1.